\magnification=1200
\overfullrule=0pt
\baselineskip=20pt
\parskip=0pt
\def\dag{\dagger}
\def\del{\partial}

\def\a{\alpha}     
\def\b{\beta}      
\def\g{\gamma}     \def\G{\mit\Gamma}
\def\d{\delta}     
\def\e{\epsilon}   
\def\z{\zeta}      
       
\def\q{\theta}

\def\l{\lambda}    
\def\m{\mu}	   
\def\n{\nu}

\def\p{\pi}        
\def\r{\rho}       
\def\s{\sigma}

\def\f{\phi}       
\def\h{\chi}       
       
\def\w{\omega}     \def\W{\mit\Omega}
\def\br{\langle}
\def\ke{\rangle}
\def\ve{\vert}

{\settabs 5 \columns
\+&&&&UdeM-GPP-TH-98-50\cr}
\bigskip
\centerline{\bf Quantum Hall Ferromagnets: }
\centerline{\bf Induced Topological term and electromagnetic interactions}
\bigskip\bigskip
\centerline{Rashmi Ray$^1$ }
\bigskip

\centerline{Laboratoire de Physique Nucl\'eaire} 
\centerline{ Universit\'e de Montr\'eal}
\centerline{Montr\'eal, Quebec H3C 3J7, Canada}
\bigskip
\centerline{\bf Abstract}
\bigskip
The $\n =1$ quantum Hall ground state in materials like GaAs is 
well known to be ferromagnetic in nature. The exchange part of the Coulomb
interaction provides the necessary attractive force to align the electronic
spins spontaneously. The gapless Goldstone modes are the angular deviations
of the magnetisation vector from its fixed ground state orientation. 
Furthermore, the system is known to support
electrically charged spin skyrmion configurations. It has
been claimed in the literature that these skyrmions are fermionic owing to
an induced topological Hopf term in the effective action governing the
Goldstone modes. However, objections have been raised against the method
by which this term has been obtained from the microscopics of the system. 
In this article, we use the technique of the derivative expansion to derive,
in an unambiguous manner, the effective action of the
angular degrees of freedom, including the Hopf term.
Furthermore, we have coupled perturbative electromagnetic fields to the
microscopic fermionic system in order to study their effect on the spin 
excitations.
We have obtained an elegant expression for the electromagnetic coupling
of the angular variables describing these spin excitations.
\vfill
\noindent{$^1$E-mail address: rray@lpshpb.lps.umontreal.ca }

\vfill\eject
\centerline{\bf I. Introduction}
\bigskip
It is well known that solitons play an important role in condensed
matter physics [1]. Moreover, in various models describing condensed matter
systems, solitons are found to 
have unexpected quantum numbers associated with them. For instance,
as shown in [2], the skyrmionic solutions of the purely bosonic 
O(3) non-linear sigma model (NLSM) can become fermionic if a topological
term (the Hopf term) is added on with the appropriate coefficient to the
action.

The NLSM actually provides a natural description of the low energy dynamics
of the Goldstone modes in a system where some continuous global
symmetry of the action is broken spontaneously by the ground state.
For instance, in superfluid $^3 He-A$, the SO(3) spin rotation group
is spontaneously broken down to the group SO(2) about some preferred
axis [3]. Since this state is also known to be 
anti-ferromagnetic, one can express the dynamics of the Goldstone
modes, which here are anti-ferromagnetic magnons, in terms of a
NLSM. What is more interesting is that a Hopf term is naturally
generated in the effective theory for the magnons [4]. The
significance of these induced Hopf terms in a variety of systems has
been studied in [5]. In the liquid Helium system, the induced
Hopf term happens to be P and T invariant, as the anti-ferromagnetic
ground state evinces these discrete symmetries. However, in suitable
systems, Hopf terms that break these discrete symmetries could be
induced dynamically. We shall discuss such a system in this article.

Recently, it has been noticed [6] that the $\nu = 1$ quantum
Hall state in samples like GaAs (where the effective gyromagnetic
ratio $g \ll 2$ ) is ferromagnetic. Thus, the global SU(2) spin
symmetry of the microscopic action governing the 2-component fermions
is broken spontaneously to U(1) by the ground state. The ferromagnetic
ground state also breaks P and T. The Goldstone modes are the
ferromagnetic magnons. An effective action governing the dynamics
of these has been obtained from the microscopics in 
[7]. It is further known that the skyrmions in this
system are electrically charged and are favoured over the quasielectron
excitations as the charge carriers. A natural question that one should
pose is whether these skyrmions are bosonic or fermionic. It has
been claimed [8] that they are fermionic due to an induced
Hopf term. However, objections have been raised [9] against
the derivation presented therein. An induced Hopf term with the 
appropriate coefficient
has also been obtained, but within an ansatz, in [10]. In this
article, we propose to address the issue of the existence of an
induced Hopf term in a direct and straightforward manner.

The basic idea that we exploit is as follows. As we know, the ground
state is ferromagnetic. Thus, the magnetisation vector is globally
aligned in a particular direction, say the $\hat z $ direction, in
the laboratory frame. In the excited state, it suffers local deviations
from this fixed direction. We can however transform the system to a
frame such that the magnetisation is locally in the $\hat z $ direction. 
The price that we pay for this is that the fermions are now coupled 
minimally to pure SU(2) valued gauge potentials (with zero curvature) 
arising from the
transformation of coordinates to the local frame. We can then
integrate the fermions out to obtain an effective action for the
SU(2) potentials. Since these potentials encapsule all the information
of the angular deviation of the magnetisation in the excited state,
this effective action is also the effective action governing the
spin excitations. The functional determinant obtained upon 
integrating the fermions out can be expanded within a derivative
expansion scheme to obtain the required low energy effective
action [17].

The spin skyrmions in this system are electrically charged. Thus it
would be very interesting to investigate the manner in which the 
angular variables (Goldstone modes) couple to electromagnetism.
The magnons are themselves neutral. Hence we should not expect
minimal coupling [15]. In the sequel, we investigate the form of the
non-minimal coupling by subjecting the microscopic fermions to
perturbing electromagnetic fields.

The article is organised as follows. In section 2, we establish
our notation and discuss the microscopic fermionic theory that we
choose as our point of departure. In section 3, we give the form
of the effective action upon collecting all the appropriate terms
arising from the derivative expansion. Section 4 is devoted to
our conclusions and to discussions of further possibilities of
investigations along similar lines. Details of the computation have
been relegated to the appendices.
\bigskip
\centerline{\bf II. Notation and Formulation}   

\bigskip
The microscopic model for the quantum Hall system in samples like
GaAs may be taken to be:
$$
S = S_1 + S_2 \eqno (2.1)
$$
with
$$
S_1 \equiv \int dt \ \int d\vec{x}\  \psi^{\dag }(\vec x,t)
\bigl[ i\del_{t}-{1\over{2m}}(\vec p - \vec A)^2 + \mu \bigr]
\psi(\vec x,t)\eqno (2.2)
$$
$$
S_2 \equiv -{{V_0}\over 2}\int dt \ \int d\vec{x}\  
\bigl( \psi^{\dag } \psi \bigr)^2
\eqno (2.3)
$$
where $\del_{x}A^{y}-\del_{y}A^{x} = -B$, $B$ being the external
magnetic field and where $\mu $ is the chemical potential.
$\psi (\vec x,t)$ is a 2-component spinor satisfying the usual
anticommutator 
$$ \{ \psi_{\alpha }(\vec x,t), \psi^{\dag }_{\beta }
(\vec y,t)\} = \delta_{\alpha \beta }\delta (\vec x - \vec y ).
$$

Here we have taken the limit $g\rightarrow 0$ and set the Pauli term
in the action equal to zero. In this limit, the above action has an
exact spin SU(2) symmetry. If restored, a small Pauli term leads to
a soft explicit breaking of this symmetry.

More importantly, we note that in the above action, we have replaced 
the non-local repulsive Coulomb term by a local repulsive 
four-fermi interaction. While this has been done to simplify the
analysis, it is also true that the interesting physics associated
with the spontaneous symmetry breaking (SSB) is due to the
short-distance part of the Coulomb interaction [11]. The long-
distance part can be added as a perturbation.

The partition function of the system is written as
$$
Z = \int D \psi \int D \bar{\psi }\ e^{i(S_1 + S_2)}.\eqno (2.4)
$$

At this point we may use the standard property of the properly
normalised SU(2) generators, 
$t^{a}_{\a \b }t^{a}_{\g \d }={1\over 2}\d_{\a \d }\d_{\b \g }
-{1\over 4}\d_{\a \b }\d_{\g \d }$
to re-organise the Coulomb term as [12]
$$
S_2 = V_0\int dt \int d\vec x\ \bigl[ {\vec S}^2 + {1\over 4}{\r }^2
\bigr] \eqno (2.5)
$$
where $\r \equiv \bar \psi \psi $ is the density and
$S^{a} \equiv \bar \psi t^{a} \psi $ is the spin density. Here,
$t^{a}\equiv {{{\sigma }^{a}}\over 2}$, where the $\sigma^{a}$ are the
standard Pauli matrices.

Written in this form, it is quite clear the the contact interactions
in (2.5) are attractive. This in turn permits us to perform the
Hubbard-Stratonovich transformation to rewrite the quartic
fermionic terms as bilinears at the cost of introducing auxiliary
fields.

Upon introducing the auxiliary fields $\vec h \equiv  h \hat{n} $, where
$\hat{n}$ is a unit vector and $\phi $, the partition function is written as
$$
Z = \int D\phi D\vec h\  e^{-{{i}\over {V_0}}\int dtd\vec x\  [{\phi }^2 
+{1\over 4}h^2]}\int D\psi D\bar{\psi }\ e^{i\int dt d\vec x\  \bar{\psi }
[i\del_{t}+\mu +\phi -{1\over{2m}}(\vec p - \vec A )^2 + {{h}\over 2}
\hat{n}\cdot \vec {\sigma }]\psi }. \eqno (2.6)
$$

Assuming that the $\phi $ integral and the $h$ integral have non-zero
saddle points, the gap equations yielding these saddle points are

$$
\phi_0 = -{{iV_0}\over 2}\ {\rm tr}\  \br \vec x,t\vert {1\over{i\del_{t}+\mu 
+\phi_0 -{1\over{2m}}(\vec p -\vec A)^2 + {{\l }\over 2}\s_{z}}}\vert \vec x,t
\ke \eqno (2.7)
$$
and
$$ 
h^{a}_0 = -2iV_0\ {\rm tr}\  \br \vec x,t \ve t^{a}{1\over{i\del_{t}+\mu +\phi_0 
-{1\over{2m}}(\vec p-\vec A)^2 + {{\l }\over 2} \s_{z}}}\ve \vec x,t \ke
.\eqno (2.8)
$$
Here $\vec h_0 \equiv \l \hat{z}$.

These may be solved to yield
$\phi_0 \simeq {{V_0}\over 2}({{B}\over {2\p }})$ and 
$h^{a}_0 \simeq V_0({{B}\over {2\p }})\delta^{a,z}$. This just means that
the ground state is of uniform density $\r_0 = {{B}\over{2\p }}$ and
uniform magnetisation $\ve \vec h_0 \ve = V_0 \r_0 $, taken to be along
the $\hat z $ direction. This is SSB. 
The fluctuations in $\phi $ and in $h$ are obviously massive while 
Goldstone's theorem guarantees that the fluctuations in $\hat n$, the
Goldstone modes, are gapless.
Upon evaluating the $\phi $ and the $h$ integrals at the saddle points, the
partition function is written as

$$
Z = \int D\hat{n} \int D\psi D\bar{\psi }\ e^{i\int dtd\vec x\  \bar{\psi }
[ i\del_{t}+\mu -{1\over{2m}}(\vec p-\vec A)^2 + \z \hat{n}\cdot \s ]\psi }
\eqno (2.9)
$$

where $\z \equiv {{\r_0 V_0}\over 2}$.

As mentioned in the introduction, we may now transform the fermionic
fields unitarily such that locally they represent spin-up fermions.
Namely we introduce the space-time dependent unitary matrix $U \in SU(2)$
such that
$$
U^{\dag }\vec \s \cdot \hat{n}U = \s_{z} . \eqno (2.10)
$$
We next define the spin-up fermionic field as
$$
\h \equiv U^{\dag }\psi 
$$
and
$$
\bar \h \equiv  \bar {\psi } U .
$$

Let us further introduce the SU(2)-valued pure gauge potentials
$$
\W^{a}_{\m }t^{a}\equiv  U^{\dag }i\del_{\m }U .\eqno (2.11)
$$
Since these are pure gauge configurations the corresponding field strengths
vanish. Thus we have the relation [4]
$$
F^{a}_{\m \n }\equiv  \del_{\m }\W^{a}_{\n }-\del_{\n }\W^{a}_{\m }
+ \e^{abc}\W^{b}_{\m }\W^{c}_{\n } = 0. \eqno (2.12)
$$

In terms of this SU(2)-valued connection, the partition function is
written as
$$
Z = \int D\hat{n} \int D\h D\bar{\h }\ e^{i\int dt d\vec x\  \bar{\h }
[i\del_{t}+\m + \W^{a}_0 t^{a} - {1\over{2m}} (-i\del_{i} - A^{i} - \W^{a}_{i}
t^{a})^2 + \z \s_{z}]\h }. \eqno (2.13)
$$

Integrating the fermionic fields out, we can write the partition function
as a path integral over the gapless angular degrees of freedom.
Namely,

$$
Z = \int D\hat{n} \ e^{iS_{eff}}
$$

where

$$
S_{eff}=-i\ {\rm tr} \ln \bigl[ i\del_{t}+\m + \W^{a}_0 t^{a} +\z \s_{z}
-{1\over{2m}}(-i\del_{i}-A^{i}-\W^{a}_{i}t^{a})^2 \bigr]. \eqno (2.14)
$$

If we had coupled perturbative slowly varying electromagnetic fields
minimally to the microscopic fermions, the effective action would be

$$
S_{eff}=-i\ {\rm tr} \ln \bigl[ i\del_{t}+\m - a_0 + \W^{a}_0 t^{a} + \z \s_{z}
-{1\over {2m}}(-i\del_{i}-A^{i}-a^{i}-\W^{a}_{i})^2 \bigr], \eqno (2.15)
$$

where $a^{\m }$ are the slowly varying perturbing electromagnetic potentials.

We can write the operator-valued argument, $\hat O$ of the above functional 
determinant as

$$
\hat O \equiv  i\del_{t}+\m -h_0 -V \eqno (2.16)
$$

where $h_0$ is the part that can be diagonalised readily and $V$ is the
perturbation.

Here,

$$
h_0 \equiv  {1\over{2m}}(\vec p - \vec A)^2 - \z \s_{z}. \eqno (2.17)
$$

We define $\p^{i} \equiv  -i\del_{i} - A^{i}$. Making the holomorphic and the
anti-holomorphic combinations:
$\p \equiv  \p^{x}-i\p^{y}$ and $\p^{\dag }\equiv  \p^{x}+i\p^{y}$, with
$$
[\p , \p^{\dag }] = 2B \eqno (2.18)
$$
we can rewrite 
$$
h_0 = {1\over{2m}}(\p^{\dag }\p + B) - \z \s_{z}. \eqno (2.19)
$$
The spectrum of this operator is infinitely degenerate (${{B}\over{2\p }}$
states per unit area) and this degeneracy is exposed in terms of the
so-called ``guiding-centre" coordinates 
$X \equiv  x - {1\over{B}}\p^{y}$ and $Y \equiv  y + {1\over{B}}\p^{x}$. 
We form
the combinations 
$Z \equiv  X+iY$ and $\bar Z \equiv  X-iY$ with the commutation relation
$$
[Z,\bar Z]={2\over{B}}. \eqno (2.20)
$$

We see that $X$ (or $Y$) commutes with $h_0$. Thus an eigenbasis for
$h_0$ is chosen to be $\{ \ve n,X,\a \ke \} $
with $n=0,1,2,\dots \infty $, $-\infty \leq X \leq \infty $ and $\a = \pm 1$.
The index $n$ denotes a Landau level (L.L.) and $\a $ denotes the spin 
(whether ``up" or ``down").

$\p $ is the lowering operator and $\p^{\dag }$ the raising operator
for the L.L. index.
Namely,
$$
\p \ve n,X,\a \ke = \sqrt{2Bn}\ve n-1,X,\a \ke \eqno (2.21)
$$
and
$$
\p^{\dag }\ve n,X,\a \ke = \sqrt{2B(n+1)}\ve n+1,X,\a \ke .\eqno (2.22)
$$
Further,
$$
\hat X \ve n,X,\a \ke = X \ve n,X,\a \ke \eqno (2.23)
$$
and
$$
\s_{z}\ve n,X,\a \ke = \a \ve n,X,\a \ke . \eqno (2.24)
$$
Thus,
$$
h_0 \ve n,X,\a \ke = [(n+{1\over 2})\w_{c} - \z \a ]\ve n,X,\a \ke 
\eqno (2.25)
$$
where $\w_{c}\equiv  {{B}\over {m}}$ is the cyclotron frequency.

From (2.25), it is clear that the gap between opposite spins for $n=0$
is given by $2\z $. Thus the inter-electron interaction provides
an effective Zeeman gap between opposite spins. In the following, we assume
that $\z \equiv  {{\r_0 V_0}\over 2} \ll \w_{c}$.

The ferromagnetic many-body ground state is constructed out of these
single particle states by filling up all the degenerate states
with $n=0, \a =1$. This precise definition of the ground state is crucial 
in defining the functional determinant that we have obtained upon integrating
out the fermions.

From (2.15) and (2.16), we see that
$$
S_{eff}=-i\  {\rm tr} \ln [i\del_{t}+\m - h_0 - V]. \eqno (2.26)
$$
Upon expanding the logarithm, we get
$$
S_{eff}=-i\ {\rm tr} \ln [i\del_{t}+\m -h_0] + i\ {\rm tr} \sum_{l=1}^{\infty }
{1\over{l}} \bigl( G V \bigr)^{l} \eqno (2.27)
$$
where $[i\del_{t}+\m -h_0]G = I$. 

Now let us define $\hat p_0$ such that $[\hat t, \hat p_0 ] = -i$
with $\br t \ve \hat p_0 = i\del_{t} \br t \ve $. Let us introduce the
basis $\{ \ve \w \ke \} $ with $\hat p_0 \ve \w \ke = \w \ve \w \ke $.
Furthermore, $ \br \w \ve t \ke = {1\over{\sqrt{2\p }}}\ e^{i\w t}$.

We also introduce the spin projection operators 
$ P_{+}\equiv  {1\over 2}(I + \s_{z})$ and 
$ P_{-}\equiv  {1\over 2}(I - \s_{z})$, which project onto $\a = \pm 1$
respectively.

Now, $\{ \ve n,X,\w \ke \}$ is a basis that diagonalises $ p_0 +\m -h_0 $
and consequently, $G$.

Let
$$
G \ve n,X,\w \ke \equiv {\G }^{(n)}(\w ) \ve n,X,\w \ke . \eqno (2.28)
$$
Now, the Green's function $G$ can be related to the mean ground state density
through the standard relation $ \r (\vec x,t) = -i \lim_{\delta \rightarrow 
0_{+}} G(\vec x,t;\vec x,t+\delta )$. In this case we know that the mean
density gets contributions from only the $n=0$ single particle states.
This in turn tells us that
$$
\G^{(0)}(\w ) P_{+} = {1\over{\w +\m - {{\w_{c}}\over 2}+\z -i\e }}P_{+}
\eqno (2.29)
$$
$$
\G^{(0)}(\w ) P_{-} = {1\over{\w +\m - {{\w_{c}}\over 2}-\z +i\e }}P_{-}
\eqno (2.30)
$$
and for $n \neq 0$,
$$
\G^{(n)}(\w ) P_{\pm } = {1\over{\w +\m -(n+{1\over 2})\w_{c} \pm \z + i\e }}
P_{\pm } . \eqno (2.31)
$$

Henceforth, we choose $\m = {{\w_{c}}\over 2}-\z $.

Having specified the pole structure of the Green's functions, let us now
focus our attention upon the perturbation $V$. We recall that the perturbations
are functions of the coordinate operators $\hat x$ and $\hat y$, or
alternatively of $\hat z \equiv \hat x + i \hat y$ and 
$ \hat{\bar z} \equiv  \hat x - i \hat y$. In the L.L. basis, it is convenient
to write:
$\hat z = \hat Z - {{i}\over{B}}{\hat \p }^{\dag }$ and
$\hat {\bar z} = \hat {\bar Z} + {{i}\over{B}}\hat \p $.
Thus, a function of the coordinate operators may be Taylor expanded around
$\hat Z$ and $\hat {\bar Z} $ as
$$
f(\hat z, \hat {\bar z})=\sum_{p,q} {1\over{p!\ q!}}
(-{{i}\over{{B}}})^{p}({{i}\over{{B}}})^{q}
(\hat {\p^{\dag }})^{p}(\hat \p )^{q} \sharp \del^{p}_{Z} \del^{q}_{\bar Z}
f(\hat Z, \hat{\bar Z})\sharp 
$$
where $\sharp \cdots \sharp $ just indicates that the normal ordering with
respect to $\p $ and $\p^{\dag }$ forces the $Z$ and $\bar Z$ to be
anti-normal ordered. We further note that $\p ,\p^{\dag } \sim \sqrt{B}$.
Thus this Taylor expansion is also an expansion in inverse powers of $B$.
Therefore, for a large value of $B$ ($\sim 10$ T), the higher derivative terms
should become more and more marginal.

Let us now define
$$
{\cal A}_0 \equiv  a_0 - \W^{a}_0 \eqno (2.32)
$$
and
$$
{\cal A}^{i} \equiv  a^{i} + \W^{a}_{i} . \eqno (2.33)
$$
In terms of these, the perturbation is written as
$$
V={\cal A}^0 + {1\over{2m}}{\cal A}^{i}{\cal A}^{i} - {{\cal B}\over {2m}}
- {1\over{2m}}(A\p + \p^{\dag }\bar A) \eqno (2.34)
$$
where $A \equiv  {\cal A}^{x}+i{\cal A}^{y}$, $\bar A \equiv  {\cal A}^{x}-i
{\cal A}^{y}$ and ${\cal B}\equiv \del_{x}{\cal A}^{y} - \del_{y}{\cal A}^{x}$.

Using the aforementioned Taylor expansion, we may write
$$
V=V^{({1\over 2})}+V^{(0)}+V^{(-{1\over 2})}+\cdots \eqno (2.35)
$$
where the ellipses indicate terms subleading in $B$. 
Here,
$$
V^{({1\over 2})}=-{1\over {2m}}\bigl( \sharp A \sharp \p + 
\p^{\dag }\sharp \bar A
\sharp \bigr) \eqno (2.36)
$$
$$
V^{(0)}=\sharp \bigl[{\cal A}^0 - {1\over {2m}}{\cal B} 
+ {1\over{2m}}({\cal A}^{i}
)^2 \bigr] \sharp  \eqno (2.37)
$$
and
$$
V^{-({1\over 2})}={{i}\over{B}}\bigl[ \sharp \del_{\bar Z}{\cal A}^0 \sharp \p 
- \p^{\dag }\sharp \del_{Z}{\cal A}^0 \sharp \bigr]. \eqno (2.38)
$$
\bigskip
\centerline{\bf III. The Effective Action for the Goldstone Modes.}
\bigskip
In this section, we shall compute the effective action given in
(2.27) to $O(1/B)$. The first term in (2.27) does not contain the
gauge fields and is not interesting for our purposes.
To the required order, we need to compute
$$
S_{eff}=i\  {\rm tr} \bigl[ GV^{(0)} + {1\over 2}GV^{({1\over 2})}G
V^{({1\over 2})} + GV^{({1\over 2})}GV^{(-{1\over 2})}
+ GV^{({1\over 2})}GV^{({1\over 2})}GV^{(0)} \bigr] \eqno (3.1)
$$
where we have used the cyclic property of the trace.

At this point let us explain our {\it modus operandi} by providing
some concrete examples.

The simplest term is where there is only one insertion of the
perturbing potential. Namely, we focus on
$$
S^{(1)}\equiv {\rm tr}\ GV . \eqno (3.2)
$$
Upon introducing the basis where $G$ is diagonal,
$$
S^{(1)}=i\ {\rm tr}\  \sum^{\infty }_{n=0}\int dX \int d\w {\G }^{(n)}(\w )
\br n,X,\w \ve V(\hat t)\ve n,X,\w \ke . \eqno (3.3)
$$
This in turn may be written as
$$
S^{(1)}=i\ {\rm tr}\  \sum^{\infty }_{n=0}\int dt \int dX 
\int {{d\w }\over{2\p }} {\G }^{(n)}\br n,X \ve V(t)\ve n,X \ke .
\eqno (3.4)
$$
Using the fact that
$$
\int {{d\w }\over{2\p }}\ e^{i\w \delta }{\G }^{(n)}(\w )
= iP_{+} \eqno (3.5)
$$
we get
$$
S^{(1)}=-{\rm tr}\  P_{+}\int dt \int dX \br 0,X \ve V^{(0)} \ve 0,X \ke .
\eqno (3.6)
$$
It may be readily shown that (see Appendix A)
$$
\int^{\infty }_{-\infty }dX \br 0,X \ve \sharp f(\hat Z, \hat {\bar Z})\sharp 
\ve 0,X \ke = {{B}\over {2\p }}\int^{\infty }_{-\infty }dx \int^{\infty }_
{-\infty }dy\  f(x,y).
$$
Using this in conjunction with (2.37) and (3.6), we have
$$
S^{(1)}=\int dt \int d\vec x \bigl[ -\r_0 a^0 + {1\over 2}\r_0 {\W }^{z}_0
+ {{\w_{c}}\over{4\p }} + {{\w_{c}}\over {8\p }}(\del_{x}{\W }^{z}_{y}-
\del_{y}{\W }^{z}_{x}) -{{\w_{c}}\over {4\p }}{\rm tr}\ P_{+}({\cal A}^{i})^2
\bigr] \eqno (3.7)
$$
where $\r_0 \equiv {{B}\over{2\p }}$.
Let us now look at the term with two insertions of $V$.
$$
S^{(2)}\equiv {{i}\over 2}{\rm tr} \ G V G V .\eqno (3.8)
$$
Thus,
$$
S^{(2)} = {{i}\over 2}{\rm tr}\ \sum^{\infty }_{n=0} \int d\w \int dX
{\G }^{(n)}(\w )\br n,X,\w \ve V(\hat t) G(\hat p_0)V(\hat t) 
\ve n,X,\w \ke \eqno (3.9)
$$
where the trace is now over the spin indices.
Introducing a resolution of the identity in the form of
$I = \int dt \ve t \ke \br t \ve $, we have
$$
S^{(2)}={{i}\over 2}{\rm tr}\ \sum^{\infty }_{n=0}\int dX \int d\w \int dt
{\G }^{(n)}(\w )\br \w t \ve t \ke \br n,X \ve V(t) G(i\del_{t})
V(t)\br t \ve \w \ke \ve n,X \ke .\eqno (3.10)
$$
Thus 
$$
S^{(2)}={{i}\over 2}{\rm tr}\ \sum^{\infty }_{n=0}\int dt \int dX 
\int {{d\w }\over{2\p }}{\G }^{(n)}(\w )\br n,X \ve V G(\w + i\del_{t})
V \ve n,X \ke .\eqno (3.11)
$$
Expanding $G(\w + i\del_{t})$ around $\w $, we get
$$
S^{(2)}={{i}\over 2}{\rm tr}\ \sum^{\infty }_{n=0}\int dt \int dX \int 
{{d\w }\over {2\p }}{\G }^{(n)}(\w )\br n,X \ve \bigl[ 
VG(\w )V + i V \del_{\w }G(\w )\del_{t}V + \cdots  \bigr] \ve n,X \ke 
\eqno (3.12)
$$
where the ellipses indicate terms with higher time derivatives.
Let us look at the term $ \sum^{\infty }_{n=0}\int dt \int dX \int 
{{d\w }\over{2\p }} {\G }^{(n)}\br n,X \ve V^{(0)}G(\w )V^{(0)}\ve n,X \ke $
for a moment. 
This gives rise to an expression of the form:
$$
{\rm tr}\ \int dt \int dX \int {{d\w }\over{2\p }}\ e^{i\w \delta }
{1\over{\w -i\epsilon }}P_{+}\br 0,X \ve V^{(0)}{1\over{\w -2\z +i\epsilon }}
P_{-}V^{(0)}\ve 0,X \ke .
$$
Upon doing the $\w $ integration, we get
$$
{{-i}\over{2\z }}{\rm tr}\ \int dt \int dX P_{+}\br 0,X \ve V^{(0)}P_{-}V^{(0)}
\ve 0,X \ke .
$$
It can be readily shown however that all the terms with $\z $ in the 
denominator add up to zero. We shall therefore ignore these terms in what
follows.

With this, we can compute the non-zero contributions from $S^{(2)}$ using
the same techniques as were employed for $S^{(1)}$ and obtain:

$$
S^{(2)}=S^{(2a)}+S^{(2b)}+S^{(2c)} \eqno (3.13)
$$
where
$$
S^{(2a)}=\int dt \int d\vec x \bigl[ {{\w_{c}}\over{4\p }}{\rm tr}\ P_{+}
({\cal A}^{i})^2 -{{\w_{c}}\over{8\p }}(\del_{x}{\W }^{z}_{y}-
\del_{y}{\W }^{z}_{x})-{{\z }\over{8\p }}\bigl( (\vec {{\W }_{i}})^2 - 
({\W }^{z}_{i})^2 \bigr) - {{\z }\over{4\p }}\bigl( \vec{{\W }_{x}}
\times \vec{{\W }_{y}} {\bigr) }^{z}\bigr] \eqno (3.14)
$$
$$
S^{(2b)}+S^{(2c)}=\int dt \int d\vec x \bigl[ 
-{1\over{4\p }}\e^{\m \n \r }a_{\m }\del_{\n }a_{\r } + {1\over{4\p }}
\e^{\m \n \r }a_{\m }\del_{\n }{\W }^{z}_{\r }
+{3\over {16}}\vec{{\W }_0}\cdot (\vec{{\W }_{x}}\times \vec{{\W }_{y}})
\bigr] . \eqno (3.15)
$$

The details of the computation of $S^{(2a)}$ have been provided in Appendix
B.

At this point we note a few things about the contributions $S^{(1)}$ and
$S^{(2)}$. There is a gauge non-invariant term in $S^{(1)}$, namely
${\rm tr}\ P_{+}({\cal A}^{i})^2$, which cancels against an equal but opposite
contribution from $S^{(2a)}$. Furthermore, the term $\del_{x}{\W }^{z}_{y}-
\del_{y}{\W }^{z}_{x}$ also cancels between these two. Furthermore, the
external magnetic field breaks the P symmetry and the ferromagnetic
ground state breaks the T symmetry. Hence, one could almost anticipate the
emergence of an U(1) Chern-Simons (CS) term in the effective action, which is
manifest in $S^{(2)}$. Furthermore, more interestingly, there is a mixing
of the angular degrees of freedom with the electromagnetic degrees of
freedom through a CS like term, which is gauge invariant due to the
presence of the Levi-Civita tensor. This is the lowest order coupling of the
angular degrees of freedom with the electromagnetic fields. There is also
the Hopf term, $\vec{{\W }_0}\cdot (\vec{{\W }_{x}}\times \vec{{\W }_{y}})$
which has been induced in the effective action. However, owing to the
fact that ${\W }^{a}_{\m }$ is a pure gauge potential, the Hopf term receives
corrections from the term in the effective action with three insertions of
the perturbation.

Namely, we have to look at
$$
S^{(3)}\equiv -i\ {\rm tr}\ GV^{({1\over 2})}GV^{({1\over 2})}GV^{(0)}.
\eqno (3.16)
$$
This may be done straightforwardly and yields:
$$
S^{(3)}\simeq \int dt \int d\vec x \bigl[ -{1\over{16\p }}
\vec{{\W }_0}\cdot (\vec{{\W }_{x}}\times \vec{{\W }_{y}})\bigr] .
\eqno (3.17)
$$
Thus combining the terms together, one obtains
$$
\eqalignno{S_{eff} = \int dt \int d\vec x \bigl[ & {1\over 2}\r_0 {\W }^{z}_
{0} - {{\z }\over{8\p }}( (\vec{\W }_{i})^2 - ({\W }^{z}_{i})^2 ) - 
{{\z }\over{4\p }}(\vec{{\W }_{x}} \times \vec{{\W }_{y}})^{z} \cr 
& + {1\over{8\p }}\vec{{\W }_0}\cdot (\vec{{\W }_{x}}\times \vec{{\W }_{y}})
- \r_0a^{0} + {{\w_{c}}\over {4\p }}b - {1\over{4\p }}\e^{\m \n \r }
a_{\m }\del_{\n }a_{\r } \cr 
& + {1\over{4\p }} \e^{\m \n \r }a_{\m }\del_{\n }{\W }^{z}_{\r }\bigr].
& (3.18) \cr 
}
$$
Here, $b\equiv \del_{x}a^{y} - \del_{y}a^{x}$ is the perturbing magnetic
field.

As has been shown explicitly in Appendix C, we can express the unitary
matrix $U \in $ SU(2) in terms of the Euler angles, $\theta , \phi ,\h $.
Namely, we can write
$$
U=e^{-i{{\phi }\over 2}\s_{z}}\ e^{-i{{\theta }\over 2}\s_{y}}
\ e^{-i{{\h }\over 2}\s_{z}} . \eqno (3.19)
$$
Again, as $ U\s_{z}U^{\dag } = \vec{\s }\cdot {\hat n}$, the unit
vector $\hat n$ is given in terms of the Euler angles as
$$
\hat n = (\sin \theta \cos \phi , \sin \theta \sin \phi , \cos \theta ).
\eqno (3.20)
$$
Also, 
$$
(\vec{{\W }_{x}}\times \vec{{\W }_{y}})^{z}={1\over 2}\e^{ij}
\hat{n}\cdot (\del_{i}\hat{n}\times \del_{j}\hat{n}). \eqno (3.21)
$$
and
$$
(\vec{{\W }_{i}})^2 - ({\W }^{z}_{i})^2 = (\del_{i}\hat{n})^2 .
\eqno (3.22)
$$

With these, we can write the effective action as
$$\eqalignno{
S_{eff}=\int dt \int d\vec x \bigl[ & {{B}\over{4\p }}\cos \theta \del_{t}\phi 
-{{\z }\over{8\p }}(\del_{i}\hat{n})^2 - \z \r_{p} + {1\over{48\p }}
\e^{\m \n \r }\vec{{\W }_{\m }} \cdot (\vec{{\W }_{\n }} \times 
\vec{{\W }_{\r }}) \cr 
& +{1\over {4\p }}\e^{\m \n \r }a_{\m }\del_{\n }{\W }^{z}_{\r }
-\r_0 a^{0} +{{\w_{c}}\over{4\p }}b -{1\over{4\p }}\e^{\m \n \r }
a_{\m }\del_{\n }a_{\r } \bigr] . & (3.23) \cr }
$$
Here, $\r_{p}\equiv {1\over{8\p }}\epsilon^{ij} 
\hat{n}\cdot (\del_{i}\hat{n}\times 
\del_{j}\hat{n})$ is a topological density (the Pontryagin index density).
This means that $\int d\vec x \r_{p} = {\rm integer}$.

Let us look at the various terms in the effective action. The first term,
with a single time derivative is a so called Wess-Zumino term [13].
It is actually a Berry phase [7] and is ubiquitous in the path integral
representation of spin systems. The interesting point is that in this
derivation, the Berry phase has emerged through a derivative expansion
[14]. In principle higher order corrections to the
Berry phase with a higher number of time derivatives could be 
straightforwardly computed. The second term is the standard kinetic energy
term of the NLSM. Together, the first two terms tell us that the dispersion
relation of the Goldstone bosons are that of ferromagnetic magnons. 
If we set the perturbing electromagnetic field to zero momentarily,
we see that the fourth term, the Hopf term, has the appropriate coefficient
to make the solitons of the NLSM fermionic.  
When the electromagnetic
field is turned on, there is a mixing between the angular degrees of
freedom and the electromagnetic potentials. This is given by the fifth term
in (3.23), which can be rewritten in terms of the Euler angles as
$ -{1\over{4\p }}\sin \theta \e^{\m \n \r }a_{\m }\del_{\n }\theta \del_{\r }
\phi $. This provides a rather elegant expression for the electromagnetic
coupling of the angles parametrising the coset SU(2)/U(1). It is also
gratifying to note that the angle $\h $ has dropped out of the expression,
as it should, since only two angles are needed to describe the coset.
An interesting point to note is that one would not get this term by
naively covariantising the term $\r_{p}a^{0}$ which arises naturally
in this context, to $j^{\m }_{p}a_{\m }$, where 
$$
j^{\a }_{p}\equiv {1\over{8\p }}\e^{\a \b \g }\hat{n}\cdot 
(\del_{\b }\hat{n}\times \del_{\g }\hat{n})
$$
is a conserved topological current, and $\r_{p}=j^{0}_{p}$.
It is well known that skyrmions that exist as topological excitations
in the system are characterised by their winding number 
$N_{p}\equiv \int d\vec x \r_{p}$. the same winding number also gives
the electrical charge of the skyrmion ($e=1$). Thus it is natural that
the response of the system to an electrostatic potential $a^0$ should
be given by a term $\int dt \int d\vec x \r_{p}a^{0} $ in the effective
action. 
In (3.23), the electromagnetic interaction (the fifth term) can be written as
$$
S^{\rm em}_{eff} = -\int dt \int d\vec x \bigl[
\r_{p}a^{0} + {1\over{4\p }}\{ a^{x}(\del_{y}{\W }^{z}_0 - \del_0{\W }^{z}_{y})
+ a^{y}(\del_0{\W }^{z}_{x} - \del_{x}{\W }^{z}_0 )\} \bigr] .
\eqno (3.24)
$$
The first term is as expected. The second term, within parentheses does not
meet with our naive expectations. In terms of the perturbing electromagnetic
fields, this term can be expressed as
$$
S^{\rm em}_{eff}=-{1\over{4\p }}\bigl[ {\W }^{z}_0 b + {\W }^{z}_{x}e^{y}-
{\W }^{z}_{x}e^{x} \bigr] \eqno (3.25)
$$
where
$$
b\equiv \del_{x}a^{y}-\del_{y}a^{x}
$$
and
$$
e^{i}\equiv -(\del_0 a^{i} + \del_{i}a^{0}).
$$
What is quite remarkable is that this second term does not depend on the
details of the interaction $V_{0}$.
The purely electromagnetic terms are familiar from previous studies concerning
the electromagnetic effective action of spinless quantum Hall fermions 
[16,17].

From the effective action, we can readily compute the mean electromagnetic 
currents in the spin-textured (excited) state. 

Thus,
$$
\br j_{0} \ke = \r_{0} + \r_{p} -{1\over{2\p }}b
$$
$$
\br j_{x} \ke = {1\over{2\p }}e^{y} + {1\over{4\p }}(\del_{y}{\W }^{z}_{0}
-\del_{0}{\W }^{z}_{y})
$$
$$
\br j_{y} \ke = -{1\over {2\p }}e^{x} + {1\over{4\p }}(\del_{0}{\W }^{z}_{x}
- \del_{x}{\W }^{z}_{0}) . \eqno (3.26)
$$

This shows explicitly that the density in the excited state changes from
its ground state value by an amount which is a topological index density.

Given the effective action, one can also compute the mean magnetisation in the
excited state.
It is given by
$$
M^{a}(\vec x,t) \equiv \br \bar{\h } t^{a}\h \ke = {{\del {\cal L}_{eff}}\over 
{\del {\W }^{a}_{0}}} . \eqno (3.27)
$$

For instance the $z$ component of the magnetisation changes from its
value of ${1\over 2}\r_{0}$ in the ground state. It is given by
$$
M^{z}(\vec x,t)= {1\over 2}\r_{0} + {1\over 2}\r_{p} - {1\over{4\p }}b
\eqno (3.28)
$$
which is precisely equal to half the value of the mean density in the
excited state.
\bigskip
\centerline{\bf V.  Conclusions}
\bigskip
In this article, we have derived some new results and have rederived some
well-known ones on the subject of quantum Hall ferromagnets.

Previously, as in [7], the $\n =1$ system has been studied, but exclusively
within the L.L.L. projection approximation. Mixing with higher L.L., due to
the Coulomb interaction, was mainly disregarded, except in [10], where some
effects of the higher L.L. mixing were considered. However, since any 
interaction that is non-diagonal in the L.L. basis will cause mixing of the
levels, in particular electromagnetic interactions, it is important to have
a calculational technique that avoids explicit L.L.L. projection. In this
article, we have exploited precisely one such method. The fermionic operators
have not been projected onto the L.L.L. whilst the information regarding
the ground state of the system ( namely that only the L.L.L. single particle
states are filled ), has been coded into the pole structure of the Green's
function. In this method, the effects due to higher L.L. mixing appear
naturally through subleading terms in the derivative expansion scheme that
we have adopted.

In this manner, we have shown how a non-linear sigma model, describing the
spin excitations of the electrons, emerges simply in terms of the angular
variables describing the deviation of the magnetisation vector from its
ferromagnetic ground state orientation. We have shown further that a Hopf
term also emerges in terms of these same angular variables, with a coefficient
that is appropriate for turning the skyrmionic excitations in the system into
fermions. The Hopf term had also been derived in [8] but the manner of the
derivation there has met with criticism (see [9] for details). The same
reservations, however, should not exist against our rather straightforward
derivative expansion technique.

Apart from the very economical rederivation of known results listed above, we
have also investigated the obviously non-minimal electromagnetic coupling
of the spin excitations in the system [15]. To the leading order, we get a
gauge invariant ``Chern-Simons like" term which couples the angular
variables describing the spin excitations, to externally applied 
electromagnetic fields. This term clearly shows how the electromagnetic
currents in the system are affected by the spin excitations. This, we believe
is a new result. 

Within our approach, the U(1) valued electromagnetic potentials are treated
on the same footing as the SU(2) valued potentials describing the spin
dynamics. Since P and T are violated in the system, the emergence of
``Chern-Simons like' terms is only to be expected. In fact, three such terms
are obtained: A pure electromagnetic CS term, a Hopf term purely in terms of 
the angular degrees of freedom and the term described in the previous paragraph
which mixes the two.

If we set all angular excitations to zero, thereby freezing the spin degree
of freedom, we obtain the effective electromagnetic interaction of planar
polarised electrons, as has been described, for instance, in [16,17]. On the
other hand, upon setting the perturbative electromagnetic fields to zero, we
obtain the magnon effective action of [7]. Thus our work also provides an
unifying treatment of these two different systems.

As has been discussed in [16, 17] and in references therein, the gapless chiral
edge excitations are the lowest-lying excitations in the spin-polarised
Hall systems. In the Hall systems addressed by the present article, the
gapless spin waves are the lowest-lying excitations. Realistically such a
system would have a boundary and we could expect excitations at the boundary.
The relationship of these excitations to the magnons could bear investigating.
Furthermore, it would be interesting to shed light on the spin-textured
edge states beginning from the microscopics. Work in this direction is
currently in progress.
\bigskip
\centerline{\bf Acknowledgements}
\bigskip
I wish to acknowledge J. Soto for early discussions on the subject
and for a fruitful collaboration on a previous article on a similar topic.
T.H. Hansson must be thanked for suggesting that the spin collective
modes might be extracted through a unitary transformation.
I am indebted to R. Mackenzie and M. Paranjape for sharing their insights
with me and B. Sakita and V.P. Nair for their encouragement.
The work is partially supported by the N.S.E.R.C of Canada and the 
F.C.A.R of Quebec.
\vfill\eject
\bigskip
\centerline{\bf Appendix A}
\bigskip
In this appendix, we shall discuss the transformation of the integral
over the guiding centre coordinate $X$, with the L.L. index $n=0$ to an
integral over the spatial coordinates $x$ and $y$.

An expression that we repeatedly encounter in section III is
$$
\int dX \br 0,X \ve \sharp f(\hat{Z}, \hat{\bar Z}) \sharp \ve 0,X 
\ke . \eqno (A.1)
$$

In view of the fact that the normal ordered products of $\hat{\p }$ and
$\hat{\p^{\dag }}$ give zero matrix elements in the L.L.L., we can rewrite
(A.1) as
$$
\int dX \br 0,X \ve f(\hat{x}, \hat{y}) \ve 0,X \ke . \eqno (A.2)
$$

Inserting the identity in the form of $I = \int d\vec x \ve \vec x \ke 
\br \vec x \ve $ into (A.2), we get
$$
\int d\vec x\  f(x,y)\int dX \ve \br \vec x \ve 0,X \ke \ve^2 . \eqno (A.3)
$$

Now in the Landau gauge $\vec A \equiv (0,-Bx)$, the L.L.L. wavefunction is
$$
\br \vec x \ve 0,X \ke = \bigl( {{B}\over{\p }} \bigr)^{{1\over 4}} 
\bigl( {{B}\over{2\p }} \bigr)^{{1\over 2}} e^{-iBXy}\ e^{-{{B}\over 2}
(x-X)^2} . \eqno (A.4)
$$

Using (A.4) in (A.3), we get
$$
\int dX \br 0,X \ve \sharp f(\hat{Z}, \hat{\bar Z}) \sharp \ve 0,X \ke = 
{{B}\over{2\p }}\int d\vec x\  f(\vec x) . \eqno (A.5)
$$
\bigskip
\centerline{\bf Appendix B}
\bigskip
In this appendix, we shall, as an example, work out the contribution
$S^{(2a)}$ to the effective action, in reasonable detail.

From (3.12), 
$$
S^{(2a)}={{i}\over 2}{\rm tr}\ \sum^{\infty }_{n=0}\int dt \int d\vec x 
\int {{d\w }\over{2\p }}{\G }^{(n)}(\w )\br n,X \ve V^{({1\over 2})}
G(\w )V^{({1\over 2})} \ve n,X \ke . \eqno (B.1)
$$
Using the explicit form of $V^{({1\over 2})}$ given in (2.36), we note
that only $n=0$ and $n=1$ from the infinite sum over $n$ will contribute,
as $V^{({1\over 2})}$ can change the L.L. index by at most one and the
integral over $\w $ is non zero if and only if ${\G }^{(0)}(\w )$ is
involved.

Thus,
$$
S^{(2a)}={{i}\over{8m^2}}\ {\rm tr} \int dt \int d\vec x \int {{d\w }\over{2\p }}
\bigl[ {\G }^{(0)}(\w )\br 0,X \ve \p A G(\w )\bar{A}\p^{\dag } \ve 0,X \ke 
+ {\G }^{(1)}(\w )\br 1,X \ve \p^{\dag }\bar{A}G(\w )A \p \ve 1,X \ke \bigr] .
\eqno (B.2)
$$

We have already mentioned in the main body of the article that we shall drop
all terms with $\z $ appearing in the denominator as they add up to zero.
Thus, using $\p , \p^{\dag }$ on the single-particle states, using the
result (A.5) from Appendix A and invoking the cyclicity of the trace, we get,
$$
S^{(2a)}={{i\w^{2}_{c}}\over{4\p }}\int dt \int d\vec x \ {\rm tr} 
\int {{d\w }\over{2\p }}{\G }^{(0)}(\w )A{\G }^{(1)}(\w )\bar{A} .
\eqno (B.3)
$$

We know that only ${\G }^{(0)}(\w )P_{+}$ has the pole structure to give
a non zero value for the $\w $ integral. Hence,
$$
S^{(2a)}={{i\w^2_{c}}\over{4\p }}\int dt \int d\vec x \ {\rm tr} 
\int {{d\w }\over{2\p }}
 {1\over{\w -i\e }}P_{+}A\bigl[  {1\over{\w - \w_{c}
+i\e }}P_{+} + {1\over{\w -\w_{c} -2\z +i\e }}P_{-} \bigr] \bar{A} .
\eqno (B.4)
$$

Now, as $\z \ll \w_{c}$, we can expand in powers of ${{\z }\over{\w_{c}}}$
and write
$$
S^{(2a)} \simeq {{\w_{c}}\over{4\p }}\int dt \int d\vec x \bigl[ 
{\rm tr}\ P_{+}A\bar{A} - {{2\z }\over{\w_{c}}}{\rm tr}\ P_{+}A\bar{A} + 
{{2\z }\over{\w_{c}}}{\rm tr}\ P_{+}AP_{+}\bar{A} \bigr] . \eqno (B.5)
$$

Normally, we would have dropped the second term in (B.5) as it is
subdominant with respect to the first. In this case, however, the first term
cancels out against in $S^{(1)}$. Now, upon using the definitions of
$A$ and $\bar{A}$ from (2.34) and the trace relations for the generators
of SU(2), we obtain (3.14).
\bigskip
\centerline{\bf Appendix C}
\bigskip
As stated in (3.19), the unitary matrix $U \in $ SU(2) can be expressed
in terms of the Euler angles $\q ,\f ,\h $. 
Thus
$$
U=e^{-{{i\f }\over 2}\s_{z}} \ e^{-{{i\q }\over 2}\s_{y}}
\ e^{-{{i\h }\over 2}\s_{z}} . \eqno (C.1)
$$
Upon using the definition $U^{\dag }i\del_{\m }U \equiv {\W }^{a}_{\m }
{{\s^{a}}\over 2}$, we can write
$$
{\W }^{x}_{\m }=-\sin \q \cos \h \del_{\m }\f + \sin \h \del_{\m }\q 
$$
$$
{\W }^{y}_{\m }= \sin \q \sin \h \del_{\m }\f + \cos \h \del_{\m }\q 
$$ 
$$ 
{\W }^{z}_{\m }= \cos \q \del_{\m }\f + \del_{\m }\h . \eqno (C.2)
$$

From (C.2), it may be checked explicitly that
$$
\del_{\m }{\W }^{a}_{\n }-\del_{\n }{\W }^{a}_{\m } + \e^{a b c}
{\W }^{b}_{\m }{\W }^{c}_{\n } = 0 . \eqno (C.3)
$$

Again, from the relation $U\s_{z}U^{\dag }=\vec{\s }\cdot \hat{n}$, we
obtain
$$
\hat{n}=\bigl( \sin \q \cos \f , \sin \q \sin \f , \cos \q \bigr) . \eqno (C.4)
$$
Using (C.2) and (C.4), we get
$$
\e^{z a b}{\W }^{a}_{x}{\W }^{b}_{y} = {1\over 2}\e^{ij}\hat{n}\cdot 
(\del_{i}\hat{n}\times \del_{j}\hat{n}) . \eqno (C.5)
$$

Using the same equations, it is also simple to verify that
$$
({\W }^{a}_{i})^2 - ({\W }^{z}_{i})^2 = (\del_{i}\hat{n})^2 . \eqno (C.6)
$$

\bigskip
\centerline{\bf References}
\bigskip
\item{[1]}  R. Rajaraman, {\it Solitons and Instantons}, (North Holland,
Amsterdam, 1982)
\item{[2]}  F. Wilczek and A. Zee, Phys. Rev. Lett {\bf 51}, 2250, (1983)
\item{[3]}  A.M.J. Schakel, cond-mat/9805152
\item{[4]}  G.E. Volovik \& V.M. Yakovenko, J. Phys.: Condens. Matter 
{\bf 1}, 5263, (1989)
\item{[5]}  Z. Hlousek et. al., Phys. Rev. {\bf D 41}, 3773, (1990)
\item{[6]}  S.L. Sondhi et. al., Phys. Rev. {\bf B 47}, 16419, (1993)
\item{[7]}  K. Moon et. al., Phys. Rev. {\bf B 51}, 5138, (1995); W. Apel 
\& Yu.A. Bychkov, Phys. Rev. Lett. {\bf 78}, 2188, (1997); R. Ray \& J. Soto,
cond-mat/9708067
\item{[8]}  W. Apel \& Yu.A. Bychkov, Phys. Rev. Lett. {\bf 78}, 2188, (1997)
\item{[9]}  G.E. Volovik \& V.M. Yakovenko, Phys. Rev. Lett. {\bf 79}, 3791, 
(1997)
\item{[10]} R. Ray \& J. Soto, cond-mat/9708067
\item{[11]} K. Moon et. al., Phys. Rev. {\bf B 51}, 5138, (1995); R. Ray \& 
J. Soto, cond-mat/9708067
\item{[12]} K. Moon et. al., Phys. Rev. {\bf B 51}, 5138, (1995)
\item{[13]} E. Fradkin, {\it Field Theories of Condensed Matter Systems}, (
Addison-Wesley Publishing Company)
\item{[14]} R. Jackiw, Int. Jour. Mod. Phys. {\bf A3}, 285, (1988); P. de Sousa
Gerbert, Ann. Phys. {\bf 189}, 155, (1989); D. D\"usedau, Phys. Lett. {\bf 
B 205}, 312, (1988)
\item{[15]} J.M. Roman \& J. Soto, cond-mat/9709298
\item{[16]} R. Ray \& B. Sakita, Ann. Phys. {\bf 230}, 131, (1994)
\item{[17]} R. Ray \& J. Soto, Phys. Rev. {\bf B 54}, 10709, (1995)
\end